\documentclass[fleqn,10pt]{wlscirep}
\usepackage[utf8]{inputenc}
\usepackage[T1]{fontenc}

\usepackage{bm}
\usepackage{multirow}

\newcommand{\degreeCelsius}{\ensuremath{{}^\circ\mathrm{C}}}

\newcommand{\tlvec}[1]{\boldsymbol{#1}}

\DeclareMathOperator*{\argmin}{arg\,min}
\DeclareMathOperator*{\argmax}{arg\,max}

\title{%

Quantifying physical insights cooperatively with exhaustive search for Bayesian spectroscopy of X-ray photoelectron spectra

}

\author[1,*,+]{Hiroyuki~Kumazoe}
\author[2]{Kazunori~Iwamitsu}
\author[3]{Masaki~Imamura}
\author[3]{Kazutoshi~Takahashi}
\author[1]{Yoh-ichi~Mototake}
\author[4,5]{Masato~Okada}
\author[6,+]{Ichiro~Akai}

\affil[1]{Graduate School of Social Data Science, Hitotsubashi University, Kunitachi, Tokyo 186-8601, Japan}
\affil[2]{Technical Division, Kumamoto University, Kumamoto 860-8555, Japan}
\affil[3]{Synchrotron Light Application Center, Saga University, Tosu, Saga 841-0005, Japan}
\affil[4]{Department of Complexity Science and Engineering, The University of Tokyo, Chiba 277-8561, Japan}
\affil[5]{Research and Services Division of Materials Data and Integrated System, National Institute for Materials Science, Ibaraki 305-0047, Japan}
\affil[6]{Institute of Industrial Nanomaterials, Kumamoto University, Kumamoto 860-8555, Japan}

\affil[*]{h.kumazoe@r.hit-u.ac.jp}
\affil[+]{These authors contributed equally to this work.}

\begin{abstract}
We analyzed the X-ray photoemission 
spectra (XPS) of 
carbon 1s states in graphene and 
oxygen-intercalated graphene grown on SiC(0001) using 
Bayesian spectroscopy.  To realize highly accurate 
spectral decomposition of the XPS spectra, we proposed 
a framework for discovering physical constraints from 
the absence of prior quantified 
physical knowledge, in which we designed the prior 
probabilities based on the found constraints and the 
physically required conditions.  This suppresses the 
exchange of peak components during replica exchange 
Monte Carlo iterations and makes possible to decompose 
XPS in the case where a reliable structure model or a 
presumable number of components is not known.  As a 
result, we have successfully decomposed XPS of one 
monolayer (1ML), two monolayers (2ML), and quasi-freestanding 2ML
(qfs-2ML) graphene samples deposited on SiC substrates 
with the meV order precision of the binding energy, 
in which the posterior probability distributions 
of the binding energies were obtained distinguishably 
between the different components of buffer layer 
even 
though they are observed as hump and shoulder structures 
because of their overlapping. 
\end{abstract}
\begin{document}

\flushbottom
\maketitle
%
%
\thispagestyle{empty}


\section*{Introduction}\label{sec:introduction}

X-ray core-level photoemission spectroscopy (XPS) is a popular and
powerful tool for investigating the elemental composition of
materials~\cite{Hufner2003,Reinert_NewJPhys2005V7p97}.  Especially,
due to the short escape depth of photoelectrons excited by soft
X-rays, XPS has been applied to various surface and interface
analysis such as film thickness~\cite{Fadley_ProgSurfSci1984V16p275},
chemical states at surfaces or
interfaces~\cite{Himpsel_PhysRevB1988V38p6084,Riedl_JPhysDApplPhys2010V43p374009,Saha_JApplPhys2020V128p135702},
and atomic distortion at the interface~\cite{Suzuki_JElectrochemSoc2020V167p127505}.
The elemental information measured by XPS is revealed by line-shape
analysis for the obtained spectrum.  Although the regression analysis
of XPS spectra is a non-linear regression, the least squares method has been
used to minimize the fitting error until now.  With the least-squares
method, it is difficult to incorporate known physical property
information and evaluating the accuracy of the estimate is impossible.
In addition, it is not possible to obtain statistical guarantees for
the solution because the solution obtained depends on the initial
search values.  In this paper, to solve these problems, we apply
Bayesian spectroscopy to analyze the XPS spectra of one monolayer (1ML), two
monolayers (2ML), and quasi-freestanding 2ML (qfs-2ML) graphene layers
grown 
on SiC substrates and attempt to extract changes in the
chemical states of the graphene layers by chemical modification at the
interface.

High-quality, large-scale graphene can be formed by thermal decomposition
of SiC at elevated temperatures.  It is known that buffer layer 
are
formed between graphene and the SiC substrate and have an equivalent
structure to graphene, although the buffer layer 
adhere strongly to
the SiC substrate, while dangling bonds remain with the SiC substrate due
to lattice mismatch~\cite{Riedl_JPhysDApplPhys2010V43p374009}.  Riedl et al.
proposed that graphene C 1s spectra have four peaks, in addition to SiC and
graphene (Gr), two additional components S1 and
S2~\cite{Riedl_JPhysDApplPhys2010V43p374009}. 
S1 comes from the C atoms bound to one Si atom on the surface of SiC(0001)
and to three C atoms in the buffer layer. S2 comes from the remaining
sp$^2$-bonded C atoms in the buffer layer.
The carbon layer on this buffer layer 
exhibits the properties of graphene.  To modify the bonding
at the interface, various atoms, such as
H~\cite{Riedl_PhysRevLett2009V103p246804},
O~\cite{Oida_PhysRevB2010V82p041411,Antoniazzi_Carbon2020V167p746},
Ge~\cite{Emtsev_PhysRevB2011V84p125423},
Si~\cite{Xia_PhysRevB2012V85p045418},
Au~\cite{Gierz_PhysRevB2010V81p235408}
and Bi~\cite{Stohr_PhysRevB2016V94p085431} have been intercalated between
the buffer layer and SiC.  When atoms are intercalated beneath the buffer
layer, dangling bonds of Si are terminated by intercalated atoms to break
covalent bonds between the buffer layer and SiC.  Since the charge transfer
into the graphene layer is also modified by the intercalation, the charge
neutrality level can be controlled around the Dirac point artificially.
However, the transport properties are often degraded after the interface
modification.  To enhance the transport property of graphene, precise
control and characterization of the chemical states should be performed at
the intercalated interface.  The bonding at the modified interface would
differ depending on the intercalated atoms and on the treatment conditions.

In such cases, the reliable structure model for the intercalated interface
would often be lacking. Thus, an alternative method is strongly required
to decompose core-level photoemission spectra even in the case where a
reliable structure model or a presumable number of components is not
known.  However, we have been forced to analyze the lineshape with the
number of components and the constraint parameters assumed to validate a
plausible structure and the experimental setting.  Thus, there might be a
concern of containing subjective and empirical arbitrariness in the fitting
results in the XPS spectra.  In addition, if the line-shape analysis was
treated as a black-box tool due to its complexity, that would yield
incorrect results.  Therefore, a reproducible and reliable approach without
arbitrariness is required for XPS spectral analysis.

Recently, the result of Bayesian spectral deconvolution for core-level XPS
has been reported to realize automatic analysis of core-level XPS spectra
by incorporating the effective Hamiltonian into the stochastic model of
spectral deconvolution~\cite{Nagata_JPSJ2019V88p044003}.  3d core-level XPS
spectra of $\textrm{La}_2\textrm{O}_3$ and $\textrm{CeO}_2$ were well
reproduced and it was confirmed that the effective Hamiltonians selected by
model selection were in good agreement with the results obtained from a
conventional study.  Furthermore, the uncertainty of its estimated values,
which are difficult to obtain with the conventional analysis method, and
the reason why the effective Hamiltonian selected were also revealed by
spectral deconvolution based on Bayesian
inference~\cite{Nagata_JPSJ2019V88p044003}.

However, the difficulty in performing such Bayesian inference is 
in designing
the prior probability distribution. Since prior probability
distributions can restrict the range of parameters, if the constrain
condition for their parameters is known in advance, it can be incorporated
into the prior probability distribution. Otherwise, a distribution
that does not affect parameter estimation, such as a wide uniform distribution,
is used as the prior probability distribution. In that case, the
posterior probability distribution may exhibit multimodality due to exchange
among spectral components, which results in poor parameter estimation
accuracy.
However, high precision spectral analysis can be achieved even 
when scientists have no prior knowledge of the data by the following 
scenario.
They first analyze the spectral data without assuming prior knowledge. 
By reviewing the results of this analysis, they quantify physical constraints that were previously unnoticed or unquantifiable.
Utilizing such knowledge, for example, by constraining the range of regression parameters, they can achieve the analysis of spectral data with the desired accuracy.
The purpose of this study is to propose a framework for incorporating this natural flow of spectral data analysis conducted by scientists into Bayesian spectroscopy~\cite{Nagata_NeuralNetw2012V28p82,JPhysConfSer2018V1036p012022}.

In this article, core-level spectra in both pristine and oxygen-intercalated
graphenes grown in SiC(0001) have been analyzed by Bayesian
spectroscopy
with constraints on the values of the spectral parameters based on knowledge
of the physical properties.

\section*{Samples}\label{sec:samples}

Graphene layers were grown on \textit{n}-doped 6H-SiC(0001) using 
the face-to-face method~\cite{Yu_JElectronSpectroscRelatPhenom2011V184p100}, 
where two SiC substrates were placed one on top of the other with
a gap of 20~{\textmu}m using Ta foils.  After sufficient outgassing
at approximately 800~\degreeCelsius{} and annealing at
1200~\degreeCelsius{} to provide a well-ordered Si-terminated surface, 
samples were annealed at 1350~\degreeCelsius{} and
1400~\degreeCelsius{} to obtain 1ML and 2ML graphene, respectively.  
Qfs-2ML were obtained by annealing the 1ML sample for 10~min at 
550~\degreeCelsius{} in air.  All measurements were performed on the 
beamline BL13 at the SAGA Light
Source~\cite{Takahashi_JElectronSpectroscRelatPhenom2005V144p1093}.  
The core-level and valence-band photoemission spectra were measured using
a photon energy of 680 and 40~eV, respectively.  The Fermi energy and
the energy resolution were confirmed by measurements for the Fermi
level of the gold reference.  The overall energy resolutions were
estimated to be 0.69 and 0.04~eV for core-level and valence band 
measurements, respectively.

\begin{figure}
    \centering
    \includegraphics[width=0.9\columnwidth, pagebox=cropbox]{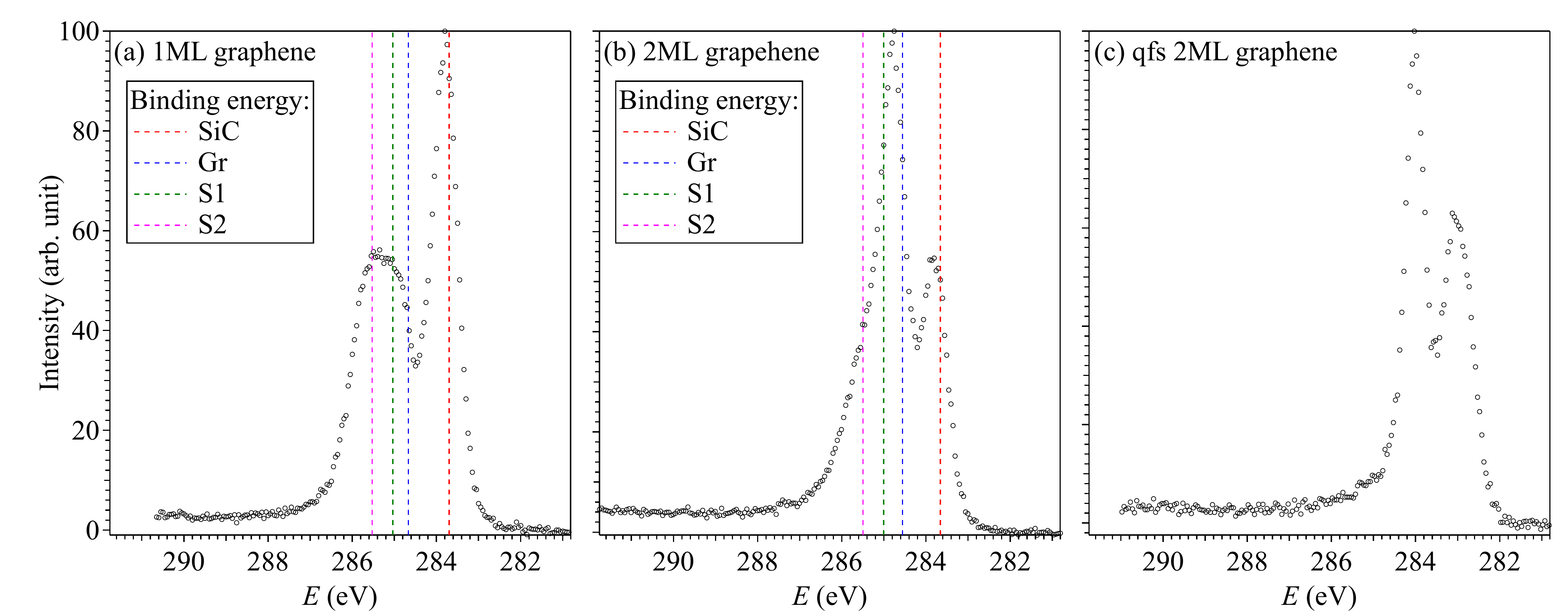}
    \caption{
        XPS spectra measured at C 1s level in 1ML~(a), 2ML~(b), and 
        qfs-2ML~(c) graphenes, respectively.  Vertical dashed lines 
        in (a) and (b) show the binding energies of SiC, graphene, 
        S1, and S2 for 1ML and 2ML graphene reported in 
        Ref.~\cite{Riedl_PhDthesis2010}.
    }
    \label{fig:Fig01}
\end{figure}
Figure~\ref{fig:Fig01} shows the XPS spectra measured
for 1ML, 2ML, and qfs-2ML graphene samples.
Vertical dashed lines 
indicate the energy positions reported in a previous 
work~\cite{Riedl_PhDthesis2010} for the 1ML and 2ML samples, which 
are shown in red for the SiC substrate, blue for graphene (Gr), 
green for S1 and magenta for S2, respectively.  Although the peak 
positions are shifted about 0.2~eV comparing the vertical dashed 
lines with the peak positions observed in Fig.~\ref{fig:Fig01}, 
this energy shift is considered to be due to differences in 
measurement conditions such as temperature and SiC doping 
concentration because the binding energy scale was calibrated with 
respect to the Fermi level of the gold reference.  

In the 1ML sample depicted in Fig.~\ref{fig:Fig01}(a), the SiC 
substrate (283.70~eV~\cite{Riedl_PhDthesis2010}) gives the strongest 
peak at 283.8~eV, and the peak structures of the graphene 
(284.67~eV~\cite{Riedl_PhDthesis2010}) and of the two components 
S1 (285.04~eV~\cite{Riedl_PhDthesis2010}) and S2 
(285.53~eV~\cite{Riedl_PhDthesis2010}) for buffer layer atoms
are observed
as broad peaks from 284.5 to 286.0~eV without separation.  In the 2ML
sample shown in Fig.~\ref{fig:Fig01}(b), the SiC substrate 
(283.66~eV~\cite{Riedl_PhDthesis2010}) gives a second intense peak at 
283.3~eV, which is the almost same as that of 1ML.  However, the
spectral structure in the high-binding energy region changes markedly, 
giving a dominant peak at 284.8~eV.  Although this peak is considered 
to be chiefly attributed to graphene (284.56~eV~\cite{Riedl_PhDthesis2010}), 
and also includes the components of the buffer layer
S1 (285.01~eV~\cite{Riedl_PhDthesis2010}) and S2 
(285.50~eV~\cite{Riedl_PhDthesis2010}) because it has a shoulder structure
on the high energy side.

The buffer layer at the interface
between SiC and graphenes consists of a carbon layer
in a graphene-like honeycomb arrangement that bonds covalently to the
Si-terminated substrate partially.  For the 1ML and 2ML samples, not all
Si atoms can bond to carbon atoms due to the different lattice constants
between SiC and graphene and due to the $30^{\circ}$ rotation angle of 
the carbon layer relative to the SiC substrate.  Although the covalent
bond breaks the hexagonal network of $\pi$ orbitals but preserves the 
$\sigma$-bonds~\cite{Riedl_JPhysDApplPhys2010V43p374009,Emtsev_MaterSciForum2007V556p525,Kim_PhysRevLett2008V100p176802}, 
it is not known whether the Si dangling bonds remain in the sample
qfs-2ML after annealing.  The XPS spectrum in the qfs-2ML sample is similar to
that in the 2ML sample, as seen in Fig.~\ref{fig:Fig01}, in which they
have two main peaks and a shoulder in the higher energy
peak.  Thus, it is considered that the two main peaks at 283.1 and 
284.0~eV come from the SiC substrate and graphene.  However, in the
qfs-2ML sample, it is controversial that there are both buffer layer components
(S1 and S2) on the shoulder of the 284.0~eV peak.  Furthermore, one can
find that the peak positions for the qfs-2ML sample are shifted to the
lower binding energy side by about 0.7~eV in Fig.~\ref{fig:Fig01}, 
although the energy axis was calibrated, and the physical reason for
this shift remains elusive.

\section*{Framework}\label{sec:method}

\begin{figure}
    \centering
    \includegraphics[width=0.9\columnwidth, pagebox=cropbox]{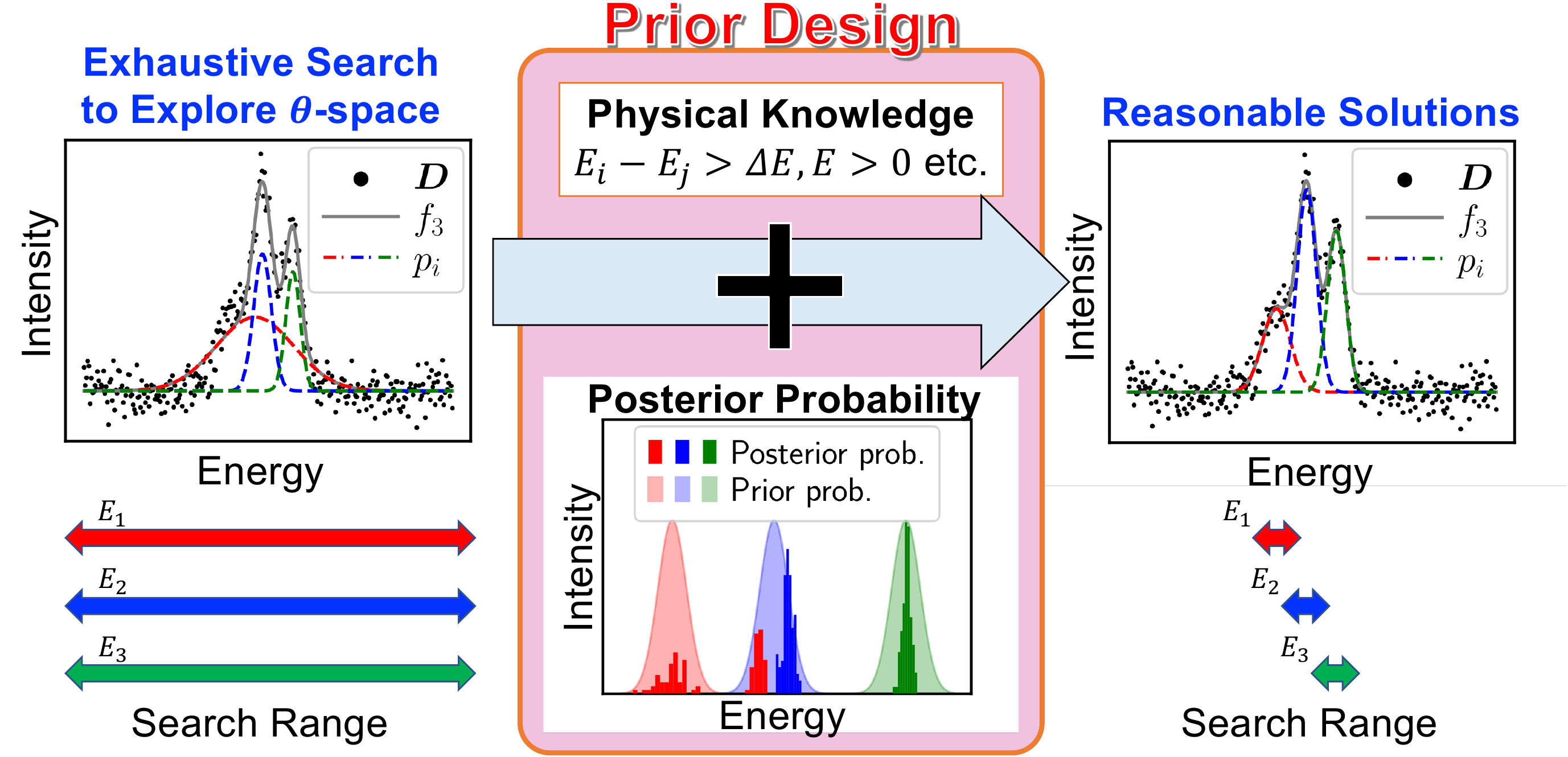}
    \caption{
        Schematic of our analysis procedure. First, we perform
        an exhaustive search to explore the parameter space.
        Next, we configure a prior probability distribution
        based on the constraints of physical properties
        and/or insights gained from the obtained posterior
        probability distribution.  Finally, we perform the
        analysis again to estimate the parameters.
    }
    \label{fig:Fig02}
\end{figure}

Figure~\ref{fig:Fig02} is the schematic diagram of our proposed
framework.  To realize the spectral decomposition with high
precision, we performed the following two-step analysis.  First, 
we performed an exhaustive search using uniform distributions
for the prior probability of parameters
$\tlvec{\theta}_{K}$ on the respective spectral components of
Gr, S1, S2, and SiC in order to explore $\tlvec{\theta}$-space.
We obtained posterior distributions of $\tlvec{\theta}_{K}$
which exhibit that the actual 
$\tlvec{\theta}$-space might be occupied by parameters.
Here, we design the prior probability distribution on 
the basis of the posterior distributions obtained by the
exhaustive search for the analysis in the next step. 
When prior information is available, we can incorporate it into
prior probability distributions.
Next, we analyzed the target data, $\tlvec{D}$ using 
the designed prior probability distribution for estimating
$\tlvec{\theta}_K$.
In Fig.~\ref{fig:Fig02}, the XPS spectrum $\tlvec{D}$ contains three peaks 
$p_i (i = 1, 2, 3)$~\cite{Tokuda_JPSJ2017V86p024001, Nagata_NeuralNetw2012V28p82}.
However, applying uniform distribution of the prior probability of
energy $E$, the width of the red component is large
and the obtained solution is not reasonable as a result of XPS
as shown in the left of Fig.~\ref{fig:Fig02}, 
and the precise estimation of the binding 
energy $E$ is inhibited because the posterior distribution of the 
red component obtained by the exhaustive search becomes 
two-modal and one distributed near the blue component as shown 
in Fig.~\ref{fig:Fig02}. Here, we design the
prior probability distribution of $E$. 
When we have constrain conditions for $E$: each
energy difference is greater than $\Delta E (> 0)$, 
we can incorporate the physical property that the red 
component is not located around the blue component.
Thus, we can configure prior probability as demonstrated in
Fig.~\ref{fig:Fig02}, and obtain reasonable solutions
with high precision.

\subsection*{Bayesian spectroscopy}\label{ssub:bayesian_spectroscopy}

Bayesian spectroscopy is a spectral decomposition analysis method
that incorporates a Bayesian inference framework.  Let 
\begin{math}
    \tlvec{D} = \{
        (x_i, y_i) \mid i = 1, \cdots, N
    \}
\end{math}
be a data set of an XPS spectrum and $f_{K}(x_i; \tlvec{\theta}_{K})$ 
be a phenomenological model function to describe $\tlvec{D}$, 
where $K$ is a subscript for model identification.  Based on $\tlvec{D}$, 
Bayesian spectroscopy evaluates the posterior probability distributions 
of the material-specific parameters $\tlvec{\theta}_{K}$ in the model 
function to be estimated.  

From Bayes' theorem~\cite{Bayes_PhilosTransRSocLond1763V53p370}, the 
posterior probability distribution $P(\tlvec{\theta}_{K}|\tlvec{D},b,K)$ 
is given by Eq.~\eqref{eq:Bayes-theorem}.
\begin{equation}
    P(\tlvec{\theta}_{K}|\tlvec{D},b,K)
    = 
    \frac{
        P(\tlvec{D}|\tlvec{\theta}_{K},b,K)
        P(\tlvec{\theta}_{K}|b,K)
    }{
        P(\tlvec{D}|b,K)
    }
    ,
    \label{eq:Bayes-theorem}
\end{equation}
where $b$ is the quasi-inverse temperature~\cite{Tokuda_JPSJ2017V86p024001} 
defined as an inverse variance $b = \sigma_{\textrm{noise}}^{-2}$ with a 
standard deviation $\sigma_{\textrm{noise}}$ of the superimposed noise in 
$\tlvec{y}=\{y_i\mid{}i=1,\cdots,N\}$.  When the noises in $\tlvec{y}$ 
are distributed independently in $i$ according to a normal distribution 
with zero mean and variance $b^{-1}$, the likelihood term 
$P(\tlvec{D}|\tlvec{\theta}_{K},b,K)$ is given by
\begin{math}
    P(\tlvec{D}|\tlvec{\theta}_{K}, b)
    =
    \left(b/2\pi\right)^{N/2}
    \exp
    \left[
    - b N \mathcal{E}_{K}(\tlvec{\theta}_{K})
    \right]
\end{math}
with an error function $\mathcal{E}_{K}(\tlvec{\theta}_{K})$ 
defined in Eq.~\eqref{eq:Error-theta}.
\begin{equation}
    \mathcal{E}_{K}(\tlvec{\theta}_{K})
    =
    \frac{1}{2N}
    \sum_{i=1}^{N} 
    \left[
        y_i 
        - 
        f_{K}(x_i; \tlvec{\theta}_{K})
    \right]^2
    .
    \label{eq:Error-theta}
\end{equation}

A Bayesian partition function 
$Z(b,K)$~\cite{Nagata_NeuralNetw2012V28p82} is obtained by 
marginalizing the numerator of Eq.~\eqref{eq:Bayes-theorem} 
over $\tlvec{\theta}_{K}$:
\begin{displaymath}
    Z(b,K)
    \equiv
    P(\tlvec{D}|b,K)
    =
    \left( 
        \frac{b}{2\pi}
    \right)^{N/2}
    \int
    \exp
    \left[ 
        - b N \mathcal{E}_{K}(\tlvec{\theta}_{K}) 
    \right]
    P(\tlvec{\theta}_{K}|b,K)
    \, d \tlvec{\theta}_{K}
    ,
\end{displaymath}
and a Bayesian free energy~\cite{Nagata_NeuralNetw2012V28p82} 
(BFE) is defined as $F(b,K)=-\ln{}Z(b,K)$.  By minimizing 
$F(b,K)$, the estimation of the noise intensity 
$\hat{\sigma}_\textrm{noise}$ superimposed on the measured 
data $\tlvec{D}$ and the model selection of the most 
appropriate function 
$f_{\hat{K}}(x_i;\tlvec{\theta}_{\hat{K}})$ to explain 
$\tlvec{D}$ can be achieved simultaneously by 
Eq.~\eqref{eq:ModelSelection}.
\begin{equation}
    \{\hat{b}, \hat{K}\}
    =
    \argmin_{b, K} F(b, K)
    ,
    \,\,
    \hat{\sigma}_\textrm{noise}
    =
    \hat{b}^{-1/2}.
    \label{eq:ModelSelection}
\end{equation}

The posterior probability distribution of material-specific 
parameters $\tlvec{\theta}_{\hat{K}}$ is sampled using a 
replica exchange Monte Carlo (RXMC)~\cite{Hukushima_JPSJ1996V65p1604} 
method according to Eq.~\eqref{eq:P-theta-D-b}.
\begin{equation}
    P(\tlvec{\theta}_{\hat{K}}|\tlvec{D},\hat{b},\hat{K})
    \propto
    \exp 
    \left[ 
        - \hat{b} N \mathcal{E}_{\hat{K}}(\tlvec{\theta}_{\hat{K}}) 
    \right]
    P(\tlvec{\theta}_{\hat{K}}|\hat{b},\hat{K})
    ,
    \label{eq:P-theta-D-b}
\end{equation}
and the maximum a posteriori probability (MAP) estimate in 
Eq.~\eqref{eq:MAP-estimation} is used for the optimal parameters 
$\hat{\tlvec{\theta}}_{\hat{K}}$ to explain the measured data 
$\tlvec{D}$.
\begin{equation}
    \hat{\tlvec{\theta}}_{\hat{K}}
    =
    \argmax_{\tlvec{\theta}_{\hat{K}}} 
    P(\tlvec{\theta}_{\hat{K}}|\tlvec{D},\hat{b},\hat{K})
    .
    \label{eq:MAP-estimation}
\end{equation}

In spectral decomposition, when there is no prior knowledge of 
material-specific parameters, its prior probability 
$P(\tlvec{\theta}_{K}|b,K)$ in Eq.~\eqref{eq:Bayes-theorem} 
should not be restricted.  However, we have to search the 
high-dimensional parameter space extensively, and 
rejections~\cite{Iwamitsu_JPSJ2021V90p104004} of candidate 
parameters prepared in Monte Carlo steps and 
exchange~\cite{Iwamitsu_JPSJ2021V90p104004} of spectral components 
during sampling become frequent, making sampling convergence 
difficult.  On the other hand, when one wants to decompose spectra 
of specific materials and quantitatively evaluate changes in 
physical properties associated with changes in the material 
interface, as is the case in this paper, we can make positive 
efforts to incorporate the knowledge of material properties into 
the prior probabilities in Bayesian spectroscopy.

\subsection*{Phenomenological model for the XPS spectrum}
\label{sub:phenomenological_model_for_the_xps_spectrum}

A phenomenological model $f_{K}(x_i;\tlvec{\theta}_{K})$ in 
Eq.~\eqref{eq:Model-function-f} is used for the spectral 
decomposition, which is a sum of peaks with a pseudo-Voigt 
function~\cite{David_JApplCryst1986V19p63} and the Shirley 
background signal~\cite{Hufner2003}:
\begin{equation}
    f_{K}(x_i; \tlvec{\theta}_{K})
    =
    \sum_{k=1}^{K} 
    p(x_i;\tlvec{\theta}_k^{\textrm{peak}})
    +
    \frac{h}{C} 
    \int_{-\infty}^{x_i}
    \sum_{k=1}^{K} 
    p(\xi;\tlvec{\theta}_k^{\textrm{peak}})
    \, 
    d\xi
    ,
    \label{eq:Model-function-f}
\end{equation}
where $K$ is the number of peaks in the XPS spectrum and %
\begin{math}
    \tlvec{\theta}_{K} = \{
        \tlvec{\theta}_{1}^{\textrm{peak}},
        \cdots,
        \tlvec{\theta}_{K}^{\textrm{peak}},
        h
    \}
    .
\end{math}
$p(x; \tlvec{\theta}^{\textrm{peak}})$ is a pseudo-Voigt 
function in Eq.~\eqref{eq:pseudo-Voigt-func}, which is a 
linear combination of Lorentzian $L(x)$ and Gaussian $G(x)$ 
shapes, with their intensity $A$, binding energy $E$, 
spectral width $w$ at full width at half maximum (FWHM) 
and a mixing ratio $\eta$.
\begin{subequations}
    \begin{equation}
        p(x; A, E, w, \eta)
        =
        \eta \cdot L(x; A, E, w)
        +
        (1 - \eta) \cdot G(x; A, E, w)
        ,
    \end{equation}
    \begin{equation}
        L(x; A, E, w)
        =
        A
        \frac{2}{\pi}
        \frac{w}{4(x - E)^2 + w^2}
        ,
    \end{equation}
    \begin{equation}
        G(x; A, E, w)
        =
        A
        \sqrt{\frac{4 \ln 2}{\pi w^2}}
        \exp \left[
            - 4 \ln 2 \left(
                \frac{x - E}{w}
            \right)^2
        \right]
        .
    \end{equation}
    \label{eq:pseudo-Voigt-func}
\end{subequations}
$C$ in Eq.~\eqref{eq:Model-function-f} is a normalization 
constant for the Shirley background given by 
$\sum_{k=1}^{K} A_k$ and $h$ is the height of the background 
signal in $x_i \to \infty$ where the intensity of the peaks 
must be zero and only the background signal remains.

\subsection*{Computational details} 
\label{sub:computational_details}

For RXMC sampling, we prepared 100 replicas with quasi-inverse 
temperatures $b_\ell$, $b_1 = 0$ and a geometric sequence 
$b_{\ell}$ for $2\le\ell\le{}100$ with $b_2=10^{-4}$ and 
$b_{100}=10$.  In all analyses in this paper, the $\hat{b}$ 
obtained in Eq.~\eqref{eq:ModelSelection} are $0.75$ -- $1.53$, 
which falls within this $b_{\ell}$ range.  On the other hand, 
$b_{2}$ should be chosen so that the state exchange of the set 
of parameters $\tlvec{\theta}_{K}$ between the replica at 
$b_{1}$ ($=0$) is guaranteed.  In this study, we set a 
sufficiently small $b_{2}$ and confirmed that the average 
exchange ratios between these replicas are more than 90~\%{} 
in all analyses.

RXMC sampling was carried out in 1,000,000 steps after a 
sufficient burn-in phase of 600,000 steps.  We used the 
auto-tuning algorithm~\cite{Iwamitsu_JPSJ2021V90p104004} 
for the step widths to achieve the mean acceptance ratio of 
70~\%.

\section*{Incorporation of physical properties into prior probability}
\label{sub:prior_probability_distribution}

In addition to model selection~\cite{Nagata_NeuralNetw2012V28p82},
the advantage of Bayesian spectroscopy is that
appropriate knowledge of physical properties can be
incorporated into the prior probability
$P(\tlvec{\theta}_{K}|b,K)$ in Eq.~\eqref{eq:Bayes-theorem}.
In XPS, broad peaks sometimes arise from multiple
components.  Therefore, to perform a well-founded physical
analysis with high precision, the restriction in the
prior probability $P(\tlvec{\theta}_{K}|b,K)$ for
$\tlvec{\theta}_{K}$ is especially effective.

\begin{table}[ht]
    \centering

    \caption{
        Prior probabilities of the binding energy $E_{k}$
        expressed in eV.
        \textsuperscript{a}
        $\mathcal{N}(\Delta{}E_\mathrm{1ML}, 0.07)$ is a
        prior probability for the energy difference
        $\Delta{}E$ ($=E_{\mathrm{S2}}-E_{\mathrm{S1}}$)
        of the binding energies between S2 and S1 components.
    }

    \begin{tabular}{cccccc}
        \toprule
        Components & 1ML & 2ML & Ref.~\cite{Riedl_PhDthesis2010} & qfs-2ML \\
        \midrule
        SiC & \multicolumn{2}{c}{$\mathcal{N}(283.78, 0.022)$} &
            283.70 & $\mathcal{N}(283.00, 0.040)$
            \\
        \rule[8pt]{0pt}{0pt}%
        Gr  & \multicolumn{2}{c}{$\mathcal{N}(284.76, 0.026)$} &
            284.67 & $\mathcal{N}(284.02, 0.0099)$
            \\
        \rule[8pt]{0pt}{0pt}%
        S1  & \multicolumn{2}{c}{$\mathcal{N}(284.92, 0.33)$}  &
            285.04 & $\mathcal{N}(284.92, 0.20)$
            \\
        \rule[8pt]{0pt}{0pt}%
        S2  & $\mathcal{N}(285.61, 0.19)$ &
            $\mathcal{N}(\Delta{}E_\mathrm{1ML}, 0.07)$\textsuperscript{a} &
            285.53 & $\mathcal{N}(284.92, 0.20)$
            \\
        \bottomrule
    \end{tabular}
    \label{tab:Tab01}

\end{table}

\subsection*{Prior probability for binding energy}
\label{sub:prior_binding_energy}

To associate each spectral component with each physical
origin while suppressing component exchange during RXMC
sampling, we set different prior probabilities for the
binding energies $E_{k}$ to distinguish the respective
components ($k=\mathrm{1:SiC, 2:Gr, 3:S1, 4:S2}$).

To accomplish this task by merging the results of the
exhaustive search and the knowledge of the previous
study, we divide the posterior probability distributions
of $E_{k}$ obtained by the exhaustive search into
four monomodal ones with reference to the previous
study~\cite{Riedl_PhDthesis2010} and evaluate the means
$\mu_{k}$ and standard deviations $\sigma_{k}$ of the
respective monomodal posterior probability distributions.
For the prior probability of $E_{k}$, we use normal
distributions $\mathcal{N}(E_{k};m_{k},s_{k})$ in
Table~\ref{tab:Tab01}, where $m_{k}$ and $s_{k}$ are
their mean and standard deviation and are determined as
$m_{k}=\mu_{k}$ and $s_{k}=5\sigma_{k}$.  Although prior
probabilities are used here with standard deviations
larger than those evaluated in the exhaustive search,
this setting avoids imposing excessive restrictions and
allows the search for the parameter space of $E_{k}$
explored in the exhaustive search.

For 1ML and 2ML samples, we use the same prior
probabilities in the respective components of SiC, Gr,
and S1 as seen in Table~\ref{tab:Tab01} since the
binding energies of these components are approximately
equal in 1ML and 2ML samples, as confirmed in
Figs.~\ref{fig:Fig01}(a) and (b).  In the case of the
2ML sample in Fig.~\ref{fig:Fig01}(b), the S2
component appears as a shoulder structure associated
with the strong components S1 and Gr.  So, although
there is a previous study~\cite{Riedl_PhDthesis2010}
showing that the binding energy of the S2 component
does not differ significantly between the 1ML and 2ML
samples, in the 2ML sample, the prior probability of the
binding energy for the S2 component is designed as
follows: we consider the difference $\Delta{}E$
($\Delta{}E=E_\textrm{S2}-E_\textrm{S1}$) in the binding
energy of the S2 component from the S1 component and
introduce a prior probability of a normal distribution
for $\Delta{}E$ as shown in Table~\ref{tab:Tab01}, where
the mean and standard deviation are
$\Delta{}E_\mathrm{1ML}$ of the 1ML sample and 0.07~eV,
respectively.

In the case of the qfs-2ML sample, although the entire
XPS spectrum in Fig.~\ref{fig:Fig01}(c) shifts to the
lower energy side than those of the 1ML and 2ML samples,
we can determine the prior probabilities of the binding
energies for the SiC and Gr components as seen in
Table~\ref{tab:Tab01} based on the exhaustive search
in the same manner.  However, the posterior probability
distributions of the binding energies for S1 and S2
obtained in the exhaustive search are broad,
and whether the S1 component remains in the qfs-2ML
sample is controversial.  Therefore, we prepare a prior
probability of the normal distribution for S1 and S2
that has the same mean value as the S1 component in
the 1ML and 2ML samples as shown in Table~\ref{tab:Tab01},
and a large standard deviation of $0.20$~eV is used to
detect the energy shift of the components S1 and S2 in
the qfs-2ML sample.

\subsection*{Prior probabilities of other parameters}
\label{sub:prior_other_parameters}

In the exhaustive search, the posterior distributions
of the spectral width $w$ are distributed in the range
of less than about 4~eV and had a mode at about 0.9~eV.
Thus, we use the same gamma distribution;
\begin{math}
    \mathcal{G}(x;\alpha,\beta)
    =
    \beta^{\alpha}
    x^{\alpha - 1}
    e^{- \beta x}
    /\Gamma(\alpha)
\end{math}
for the prior probability distribution of the nonnegative
$w$ in all components of all data, where $\Gamma(\alpha)$
is a gamma function, $\alpha=2.138$ and
$\beta=1.265~\mathrm{eV}^{-1}$, respectively.  Therefore, the
mode value of $\mathcal{G}(x;\alpha,\beta)$ is 0.9~eV and
the range of two standard deviations from the mean covers
4~eV.   Although, to decompose the shoulder structure in
the XPS spectrum of the 2ML sample [see
Fig.~\ref{fig:Fig01}(b)] into two components, the binding
energy of S2 was parameterized by $\Delta{}E$
, we also assume that
S1 and S2 have the same spectral width
($w_{\mathrm{S1}} = w_{\mathrm{S2}}$) in the 2ML sample,
since these two buffer components are expected to have
similar lineshapes.
For other parameters in Eq.~\eqref{eq:Model-function-f},
we set $A>0$, $0\le\eta\le1$, $h>0$ as prior probabilities
using uniform distributions.

\section*{Results}\label{sec:results}

\subsection*{1ML and 2ML samples} 

\newcommand{\est}[1]{\widehat{#1}}
\newcommand{\val}[1]{\ensuremath{\est{{#1}}~(\sigma_{#1})}}
\begin{table}[ht]
    \centering
    \caption{
        MAP estimates and standard deviations of
        $P(\theta|\tlvec{D}, \hat{b})$ as measures
        of the precision of the estimation of decomposed
        components in the 1ML sample.
    }
    \begin{tabular}{lccccc}
        \toprule
             & \val{A}     & \val{E} [eV]       & \val{w} [eV]      & \val{\eta}      & \val{h}  \\
        \midrule
        SiC  & 72.4 (1.4)  & 283.7905 (0.0019)  & 0.7070  (0.0048)  & 0.100 (0.031)  & \multirow{4}{*}{2.58 (0.13)} \\
        Gr   & 23.9 (2.1)  & 284.766  (0.024)   & 0.943   (0.041)   & 0.44  (0.17)   & \\
        S1   & 24.8 (2.6)  & 285.050  (0.080)   & 0.91    (0.19)    & 0.20  (0.22)   & \\
        S2   & 48.3 (2.5)  & 285.691  (0.028)   & 0.953   (0.031)   & 0.461 (0.086)  & \\
        \bottomrule
    \end{tabular}    
    \label{tab:Tab02}
\end{table}
Table~\ref{tab:Tab02} summarizes the results
of Bayesian spectroscopy for the 1ML sample,
where $\hat{\theta}$ and $\sigma_{\theta}$
are the MAP estimates and the standard deviations
of $P(\theta|\tlvec{D}, \hat{b})$ as measures
of the precision of the estimation.
\begin{figure}[ht]
    \centering
    \includegraphics[width=0.5\columnwidth, pagebox=cropbox]{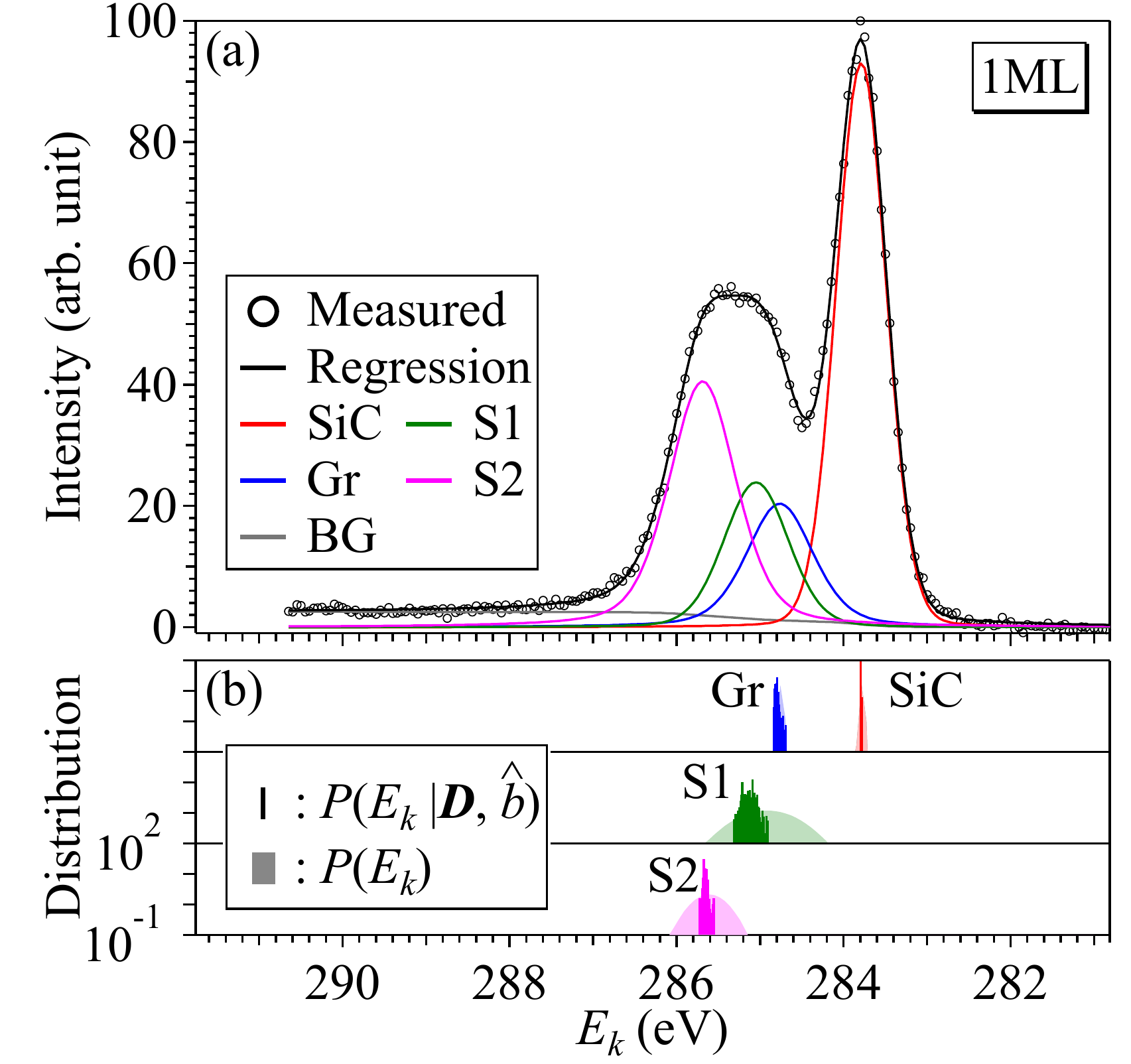}
    \caption{
        (a) Measured XPS spectrum, a regression spectrum and
        the decomposed spectral components for the 1ML sample.
        (b) Prior and posterior probability of the binding
        energy for each component.
    }
    \label{fig:Fig03}
\end{figure}
The colored and black curves in
Fig.~\ref{fig:Fig03}(a) are the spectral
components decomposed and a regression spectrum
by Eq.~\eqref{eq:Model-function-f}, and Bayesian
spectroscopy can successfully decompose the XPS
spectrum into four components and the background
signal with high reproducibility.  The root mean
square deviation (RMSD) of the regression spectrum
is 0.83 in the intensity scale of the XPS signal,
and it is consistent with the noise intensity
$\hat{\sigma}_{\mathrm{noise}}=0.86$
($=\hat{b}^{-1/2}$) estimated by the optimal
quasi-inverse temperature $\hat{b}=1.36$ in
Eq.~\eqref{eq:ModelSelection}.

Figure~\ref{fig:Fig03}(b) shows the prior and
posterior probability distributions of the binding
energy $E_{k}$, in which the light and dark colors mean
the prior and posterior ones, and the ordinate is
on a logarithmic scale.  Although the posterior
probability distribution of Gr is as broad as its
prior probability, the posterior probability
distributions become narrower in the components SiC,
S1, and S2, indicating that $E_{k}$ can be
estimated with high precision.  The probability
distributions for S1 and S2 are particularly
noteworthy in Fig.~\ref{fig:Fig03}(b).  Although
the prior probability distributions of S1 and S2
shown in light green and light magenta have overlapping
hems, the posterior probability distributions shown
in dark green and dark magenta are unimodal and noticeably
narrower with no overlap, allowing us to decompose
the hump structure into two distinguishable components
with statistical assurance.

\begin{table}[ht]
    \centering
    \caption{
        MAP estimates and standard deviations of
        decomposed components in the 2ML sample.
        \textsuperscript{a}
        we assume $w_{\mathrm{S1}} = w_{\mathrm{S2}}$ in the 2ML sample.
    }
    \begin{tabular}{lccccc}
        \toprule
            & \val{A}     & \val{E} [eV]       & \val{w} [eV]     & \val{\eta}      & \val{h}  \\
        \midrule
        SiC & 36.2 (1.3)  & 283.7832 (0.0056)  & 0.688  (0.012)   & 0.021 (0.027)  & \multirow{4}{*}{3.79 (0.13)} \\
        Gr  & 66.3 (1.6)  & 284.7601 (0.0016)  & 0.6540 (0.0064)  & 0.552 (0.083)  & \\
        S1  & 15.4 (1.2)  & 285.089  (0.064)   & 1.199  (0.056)   & 0.08  (0.10)   & \\
        S2  & 33.4 (1.1)  & 285.515  (0.035)   & N/A\textsuperscript{a}  & 0.49  (0.17)   & \\
        \bottomrule
    \end{tabular}

    \label{tab:Tab03}
\end{table}
Table~\ref{tab:Tab03} also summarizes the results
of Bayesian spectroscopy in the 2ML sample. 
The binding energy of S2 in the 2ML sample is
parameterized by the difference $\Delta{}E$ from
the binding energy of S1.  The MAP estimate of
$\Delta{}E$ is 0.426~eV and the standard deviation
of its posterior probability distribution is
0.056~eV ($<0.07~\textrm{eV}$).  The binding energy
of S2 shown in Table~\ref{tab:Tab03} is the result
obtained by sampling of $E_\textrm{S1}+\Delta{}E$.

\begin{figure}[ht]
    \centering
    \includegraphics[width=0.5\columnwidth, pagebox=cropbox]{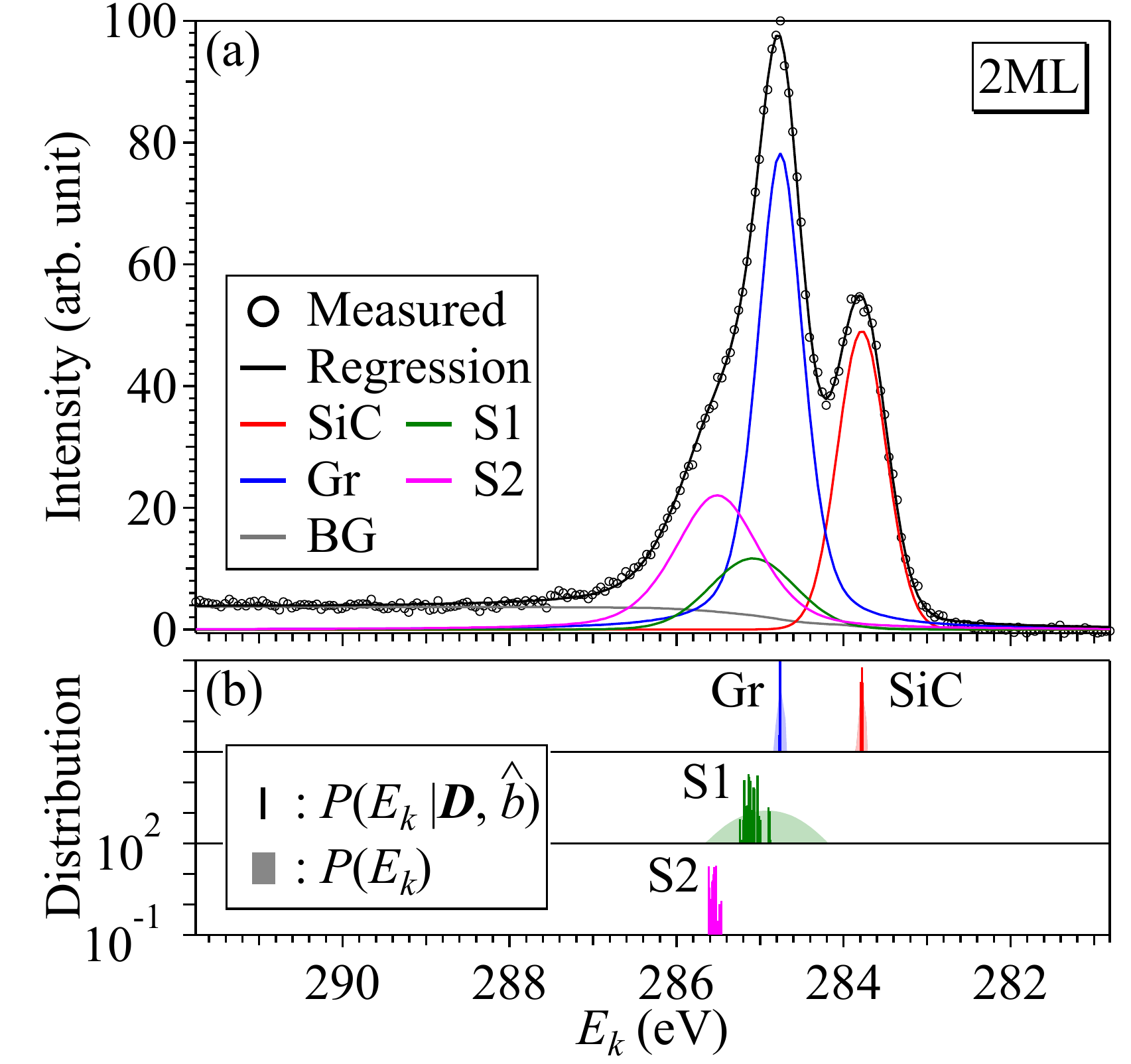}
    \caption{
        (a) Measured XPS spectrum, a regression spectrum and
        the decomposed spectral components for the 1ML sample.
        (b) Prior and posterior probability of the binding
        energy for each component.
    }
    \label{fig:Fig04}
\end{figure}
Figure~\ref{fig:Fig04}(a) also shows the results of
Bayesian spectroscopy in the 2ML sample, and the
regression spectrum indicated by a black curve can
reproduce the measured one well, and the shoulder
structure is explained by the components S1 and S2.
In contrast to the prior probabilities of the
binding energies of SiC, Gr, and S1 shown in light
colors in Fig.~\ref{fig:Fig04}(b), their posterior
probability distributions shown in dark colors are
sharp and can be estimated with high precision.
Although the prior probability distribution of
$E_\textrm{S2}$ is not shown in
Fig.~\ref{fig:Fig04}(b), because $E_\textrm{S2}$ is
parameterized by $E_\textrm{S1}$ and $\Delta{}E$
and if these two parameters are independent of each
other, the standard deviation of its prior
probability is 0.338~eV ($=\sqrt{0.33^{2}+0.07^{2}}$)
and is comparable to that (0.33~eV) of the prior
probability of $E_\textrm{S1}$.  As shown in dark
magenta in Fig.~\ref{fig:Fig04}(b), the standard
deviation of the posterior probability distribution
for $E_\textrm{S2}$ is sufficiently smaller than
its standard deviation.  Such highly precise
estimation can be achieved by incorporating the
findings of the 1ML analysis as the energy
difference between S1 and S2.

\subsection*{Qfs-2ML sample}

To estimate whether both the S1 and S2 components
are included in the shoulder structure in
Fig.~\ref{fig:Fig01}(c) for the qfs-2ML sample, we
prepare models with and without the S2 component and
perform model selection.  The BFEs of the model
with and without S2 are 357.9 and 364.4, respectively.
As a result, according to Eq.~\eqref{eq:ModelSelection},
Bayesian spectroscopy chooses the former model that
includes the S2 component.  The difference in these
BFEs is Bayesian statistically apparent, and Bayesian
spectroscopy argues that S2 is required for the XPS
spectrum of the qfs-2ML sample even after accounting
for the superimposed noise intensity
$\hat{\sigma}_\mathrm{noise}=1.15$ ($\hat{b}=0.754$).
\begin{table*}[ht]
    \centering
    \caption{
        MAP estimates and standard deviations of
        decomposed components in the qfs-2ML sample
        using the model with S2.
    }
    \begin{tabular}{lccccc}
        \toprule
            & \val{A}      & \val{E} [eV]      & \val{w} [eV]    & \val{\eta}    & \val{h}  \\
        \midrule
        SiC & 56.03 (0.51) & 283.0144 (0.0024) & 0.8730 (0.0051) & 0.002 (0.026) & \multirow{4}{*}{3.159 (0.071)} \\
        Gr  & 51.50 (0.17) & 284.0183 (0.0020) & 0.5176 (0.0045) & 0.186 (0.038) & \\
        S1  & 12.2  (4.4)  & 284.541  (0.097)  & 1.31   (0.15)   & 0.16  (0.11)  & \\
        S2  &  0.5  (4.3)  & 284.97   (0.15)   & 1.21   (0.30)   & 0.63  (0.15)  & \\
        \bottomrule
    \end{tabular}

    \label{tab:Tab04}
\end{table*}
Table~\ref{tab:Tab04} summarizes the results
of Bayesian spectroscopy in the qfs-2ML sample.
\begin{figure}[ht]
    \centering
    \includegraphics[width=0.5\columnwidth, pagebox=cropbox]{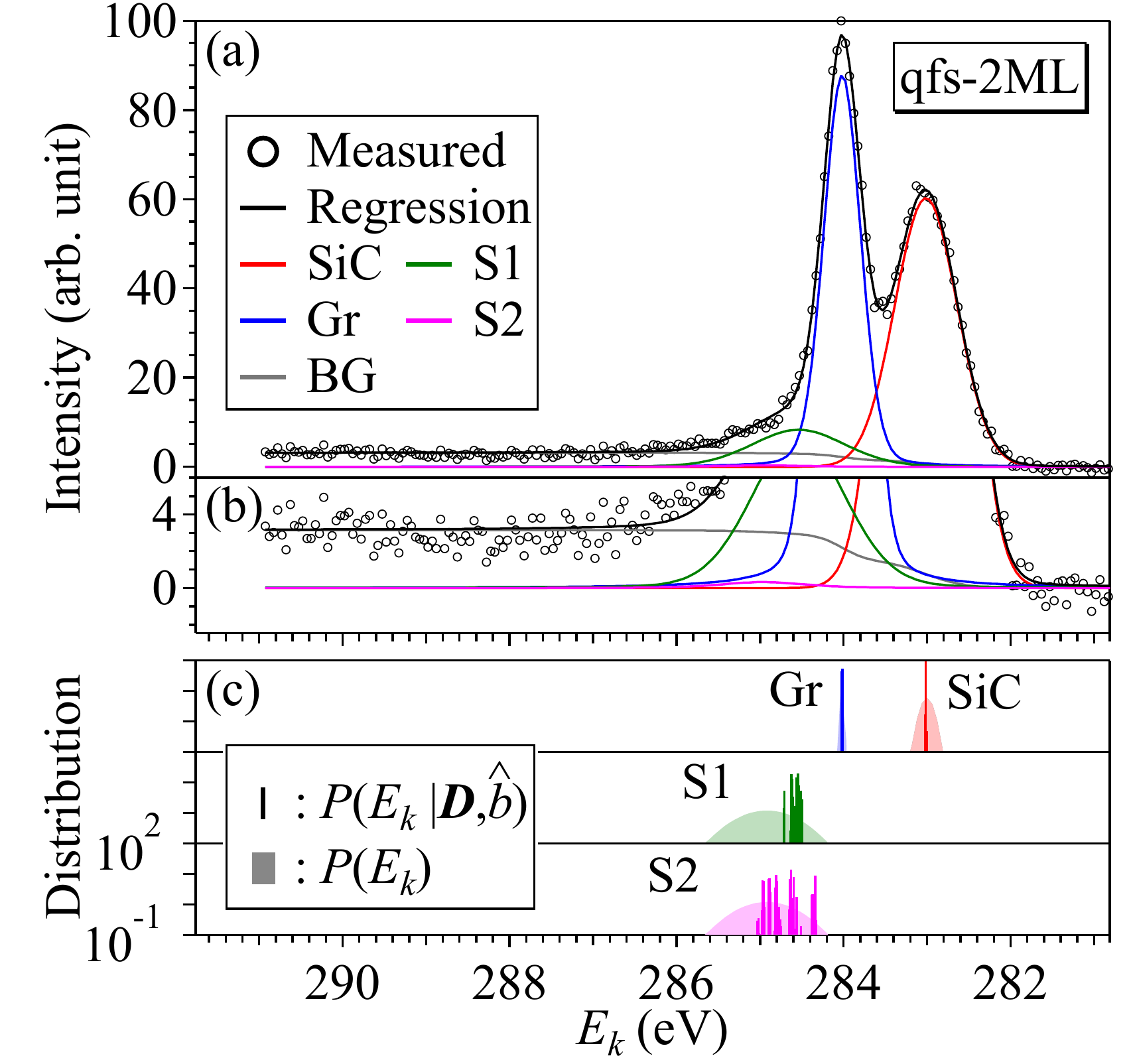}
    \caption{
        (a) Measured XPS spectrum, a regression spectrum and
        the decomposed spectral components for the qfs-2ML
        sample.  (b) Enlarged ordinate scale of (a) to show
        the weak S2 component.  (c) Prior and posterior
        probability of the binding energy for each component.
    }
    \label{fig:Fig05}
\end{figure}
The color and black curves in Fig.~\ref{fig:Fig05}(a)
are the decomposed spectral components and a regression
spectrum for the qfs-2ML sample, respectively.  The
black curve shows high reproducibility with the
measured one, and the estimated $\hat{b}$ is
consistent with the RMSD of the regression.

Figure~\ref{fig:Fig05}(b) shows on an enlarged ordinate
scale.  Although we can confirm the weak S2 component
shown in magenta, it is found that the contribution of S2
is quite small.  Bayesian spectroscopy searches for a
globally optimal solution that reproduces the entire
data using a model function, and it is possible to
extract even components whose peak intensity is lower
than the noise intensity~\cite{Iwamitsu_JPSJ2020V89p104004},
such as the S2 component in this case.
\begin{figure}
    \centering
    \includegraphics[width=0.5\columnwidth, pagebox=cropbox]{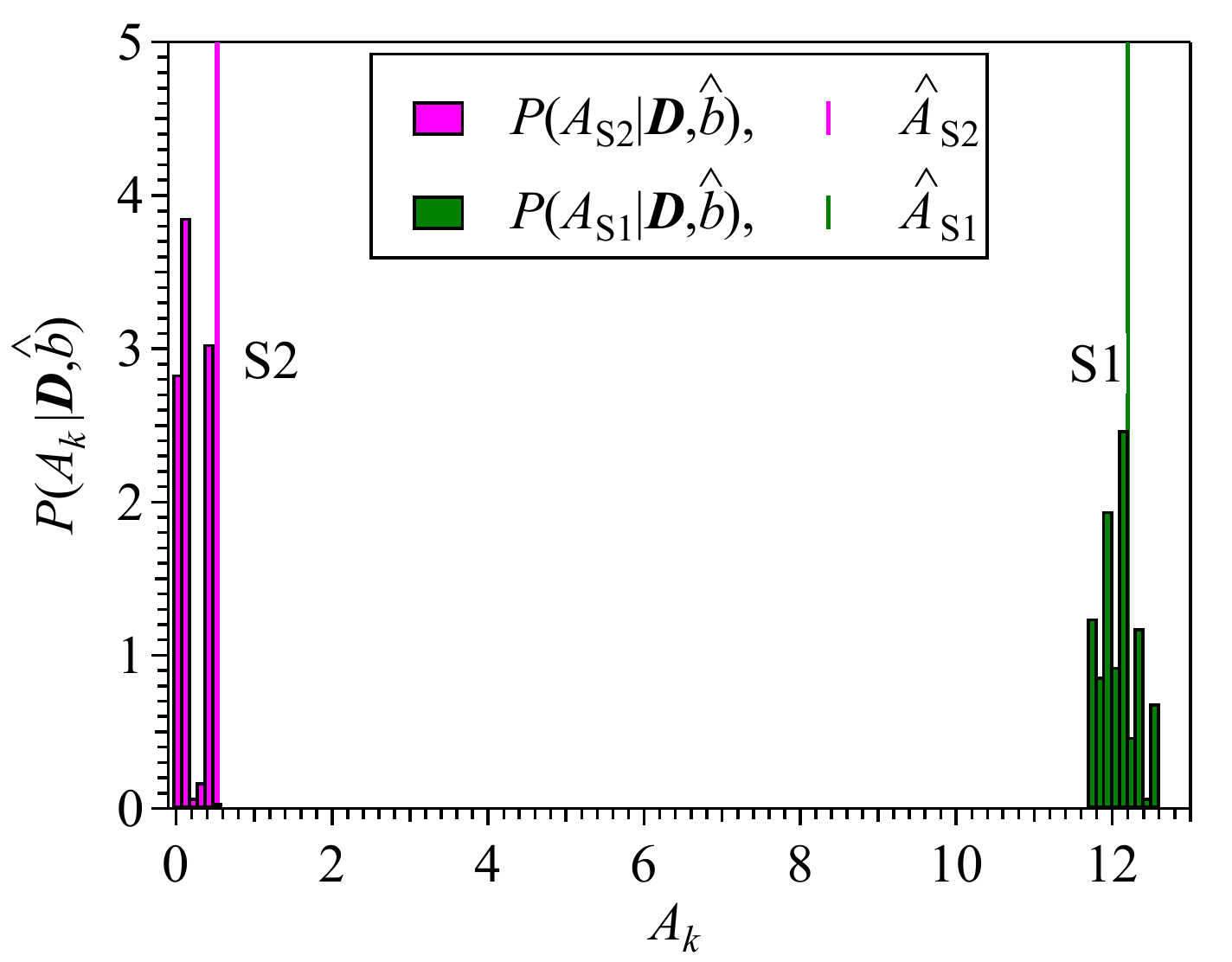}
    \caption{
    Posterior probability distributions of the integrated
    intensity for S1 and S2 components in the qfs-2ML sample.
    MAP estimates $\hat{A}_\textrm{S1}$ and $\hat{A}_\textrm{S2}$
    are indicated by vertical lines.
    }
    \label{fig:Fig06}
\end{figure}
However, we have to evaluate the integrated intensity
$A_{k}$ to discuss the presence of the buffer layer S2,
and Figure~\ref{fig:Fig06} shows the posterior probability
distributions of $A_\textrm{S1}$ and $A_\textrm{S2}$ for
components S1 and S2, in which the MAP estimates
$\hat{A}_\textrm{S1}$ and $\hat{A}_\textrm{S2}$ are
indicated by vertical lines.  Bayesian spectroscopy, in fact,
selects the model that includes the S2 component and gives
a non-zero MAP estimate ($\hat{A}_\textrm{S2} = 0.5$) as
shown in Table~\ref{tab:Tab04}.  However, the posterior
probability of $A_\textrm{S2}$ is distributed near zero
within the non-negative value constraint, and its standard
deviation is as large as $\sigma_{A_\textrm{S2}} = 4.3$
($>\hat{A}_\textrm{S2}$).  The results of this analysis
indicate that
the annealing procedure in air terminates the dangling bonds of Si atoms under the buffer layer with oxygen atoms,
causing the S2 component to disappear.
It also means that there are almost no dangling bonds
between the SiC substrate and graphene in the qfs-2ML
sample.
It is reported that an ideal oxidization proceedure can forms Si$_2$O$_5$ adlayer without dangling bond on Si-face of SiC substrate~\cite{Bernhardt_APL1999V74p1084}. All of the dangling bonds and the covalent bonding between the buffer layer and SiC substrate would be terminated if our annealing proceedure fully oxidize the SiC substrate.
$\hat{A}_\textrm{S2}$ in Table~\ref{tab:Tab04}
is approximately 1/24th of $\hat{A}_\textrm{S1}$ for the S1
component, which is considered to be the S1 component that
remained slightly after annealing.

\section*{Discussion} 
\label{sec:discussion}

According to the previous
study~\cite{Riedl_JPhysDApplPhys2010V43p374009}, the S1
component results from
the C atoms bound to Si of SiC surface and C atoms in the buffer layer, which is approximately one third of the total C atoms in the buffer layer,
and the S2 component is the result of the remaining C atoms in the buffer layer.
In a
previous study~\cite{Riedl_PhDthesis2010}, 0.31 has been
reported for the integrated intensity of the S1 component
relative to the sum of the S1 and S2 components in both
1ML and 2ML samples.  We obtain results consistent
with these previous
studies~\cite{Riedl_JPhysDApplPhys2010V43p374009,Riedl_PhDthesis2010}.
We evaluate the posterior probability distributions of
$\varLambda:=A_\textrm{S1}/(A_\textrm{S1}+A_\textrm{S2})$
from the sampling histories of $A_\textrm{S1}$ and
$A_\textrm{S2}$, and obtain MAP estimates of 0.339 for the
1ML sample and 0.315 for the 2ML sample, respectively.
The standard deviations of their posterior probability
distributions are 0.032 and 0.020, respectively, and the
values of previous studies are included within these
ranges.
\color{black}

\begin{figure}
    \centering
    \includegraphics[width=0.5\columnwidth, pagebox=cropbox]{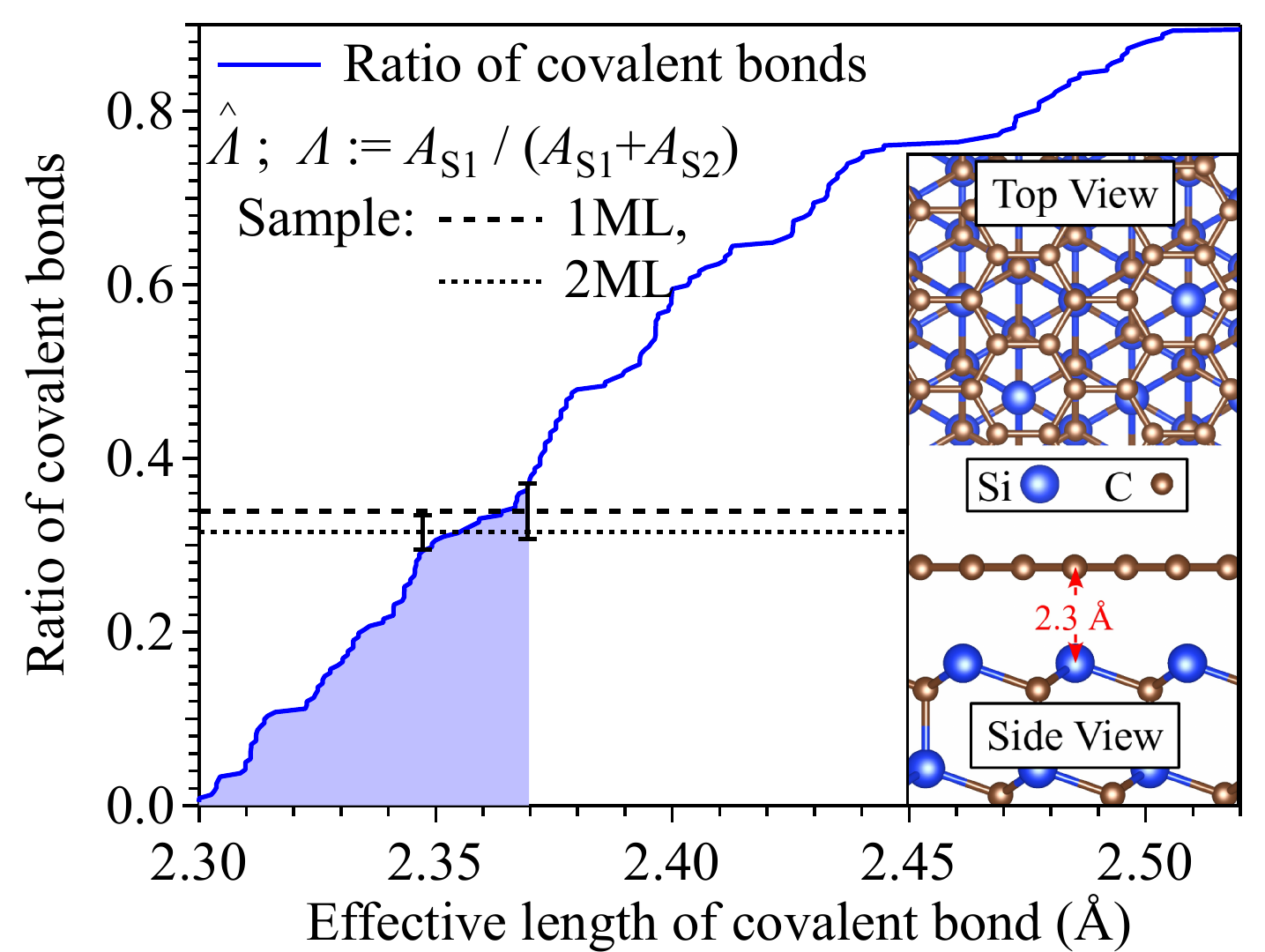}
    \caption{
        The number ratio of C covalently bonded to Si as a
        function of the effective length of its covalent bond.
        The horizontal lines and their error bars are the MAP
        estimates $\hat{\varLambda}$ and the standard
        deviations of the posterior probability distributions
        of $\varLambda$.
    }
    \label{fig:Fig07}
\end{figure}
We can estimate the effective length of the covalent bond,
which gives S1, between C and Si based on this
intensity ratio $\varLambda$, assuming that the
intensities of the XPS signal of the components S1 and S2
are equivalent and using a structural model.  A portion of
the structural model at the interface of the graphene and
SiC(0001) substrate is shown in the inset of
Fig.~\ref{fig:Fig07}.  We consider a rectangular area of
$34.045\times58.971~\textrm{\AA}^2$ on the SiC surface
tiled with $(\sqrt{3}\times\sqrt{3})~R30^{\circ}$ unit,
in which the unit cell of graphene (brown honeycomb)
rotates with $30^{\circ}$, and 242 Si atoms and 768 C atoms
are interfaced in this rectangular area.  The distance between
the SiC substrate and graphene is
2.3~{\AA}~\cite{Sumi_APL2020V117p143102}.
When the effective length of the covalent bond formation is extended to increase the number of covalent bonds with Si within the effective length, the ratio of the covalent bonds to all C atoms in the buffer layer increases as indicated by the blue curve in Fig.~\ref{fig:Fig07}. 
The horizontal lines and
their error bars are the MAP estimates $\hat{\varLambda}$
for the intensity ratio of the S1 components to the sum of S1
and S2 components and the standard deviations of the posterior
probability distributions of $\varLambda$.  Taking into
account the accuracy of the estimation of $\hat{\varLambda}$,
the measured S1 intensity ratios $\hat{\varLambda}$ can be
understood by covalently bonding to Si at distances less than
2.37~{\AA}, shown in a light blue area in Fig.~\ref{fig:Fig07}.
This is also consistent with a previous
study~\cite{Hass_PhysRevB2008V78p205424}.

The primary advantage of Bayesian spectroscopy is that it
provides estimates through statistical sampling in the
parameter space.  However, when multiple spectral components
are expected to be contained in the tail part of a strong
spectral structure and in a hump structure, as analyzed in
this paper, the results of simple statistical sampling are
deceptive.  
Our proposed method illustrated in Fig.~\ref{fig:Fig02} 
solves this problem and makes it possible to estimate 
the material-specific parameters with high accuracy.
Consequently, the
standard deviations of the posterior probability distribution
for the binding energy, shown in
Tables~\ref{tab:Tab02}--\ref{tab:Tab04} as measures of the
accuracy of the estimation of the MAP estimates, are on the
order of 100~meV even for the S2 component of the most severe
case of the qfs-2ML sample, and are less than several 10~meV
for the others, making an extremely accurate estimation of the
binding energies.

\section*{Conclusion} 
\label{sec:conclusion}

Using Bayesian spectroscopy, we have investigated the XPS 
spectra of graphene samples
in the C 1s level.  To perform the highly precise spectral 
decomposition of the XPS, we first performed an exhaustive search
to explore the parameter space and then performed 
spectral decomposition by Bayesian spectroscopy using designed 
prior probabilities that incorporate the information based 
on physical properties and the insights gained from the 
posterior probability distributions evaluated in the 
exhaustive search.  We have succeeded in decomposing the 
XPS spectra of the 1ML, 2ML, and qfs-2ML 
samples to the components: graphene, SiC, and the buffer
layer atom(s)
and estimation of the binding energies 
has been achieved with high precision of the order of meV.
From their binding energies, we estimated the existence
ratio of the buffer layers S1 for S2 with its standard 
deviation, which is consistent with previous studies.  
We performed model selection to determine the number of 
components in the XPS spectrum of the qfs-2ML sample. 
The four-peak model is selected, however, the contribution 
of S2 is quite small and this is probably considered due 
to the heterogeneity of the qfs-2ML graphene sample.
These results demonstrate that the appropriate design of 
the prior probability distributions based on the 
information of physical properties and the insights 
gained from the exhaustive search is effective to 
perform the spectral decomposition with high precision.

\bibliography{Reference}

\section*{Acknowledgements (not compulsory)}

This study was supported by JST, CREST (Grant Nos. JPMJCR1861 and JPMJCR1761), Japan; and NEDO (Grant No. JPNP22100843-0), Japan.

\section*{Author contributions statement}

All authors contributed to the study conception and design. M.I. and K.T. conducted the experiments, and H.K., K.I. and I.A. analyzed the results. H.K. wrote the first draft of the manuscript, and K.T., Y.M., M.O. and I.A. supervised the project. All authors read and approved the final manuscript.

\section*{Data availability statement}
The data that support the findings of this study are available from the corresponding author upon reasonable request.

\section*{Additional information}



\section*{Competing interests}

The authors declare no competing interests.

\end{document}